\documentclass[pre,aps,superscriptaddress,showpacs]{revtex4-1}
\usepackage{natbib}
\usepackage{amsmath}
\usepackage{epsfig}
 \usepackage{color}

\DeclareMathOperator*{\sumint}{%
\mathchoice%
{\ooalign{$\displaystyle\sum$\cr\hidewidth$\displaystyle\int$\hidewidth\cr}}
{\ooalign{\raisebox{.14\height}{\scalebox{.7}{$\textstyle\sum$}}\cr\hidewidth$\textstyle\int$\hidewidth\cr}}
{\ooalign{\raisebox{.2\height}{\scalebox{.6}{$\scriptstyle\sum $}}\cr$\scriptstyle\int$\cr}}
{\ooalign{\raisebox{.2\height}{\scalebox{.6}{$\scriptstyle\sum$}}\cr$\scriptstyle\int$\cr}}
}

\begin{document}

\title{Auxiliary open quantum system for the Floquet quantum master equation}
\author{Fei Liu}
\email[Email address: ]{feiliu@buaa.edu.cn}
\affiliation{School of Physics, Beihang University, Beijing 100191, China}

\date{\today}

\begin{abstract}
{By directly using the probability formulas of quantum trajectories, we construct an auxiliary open quantum system for
a periodically driven open quantum system whose dynamics is governed by
the Floquet quantum master equation. This auxiliary system can generate a quantum trajectory ensemble that is consistent with the canonical quantum trajectory ensemble. We find that, at a long time limit, though the Lindblad operators are modified, the coherent dynamics of the auxiliary system is the same as that of the original system. A periodically driven two-level quantum system is used to illustrate this construction. }
\end{abstract}
\maketitle

\section{Introduction}
\label{section1}
In the past decade, microcanonical and canonical trajectory ensembles of stochastic systems have attracted considerable interest~\cite{Evans2004,Garrahan2007,Maes2008,Maes2008a,Garrahan2009,Garrahan2010,Jack2010,Takahiro2011,
chetrite2013,Chetrite2014,Derrida2019}. The underlying reason is that the rare events or rare large fluctuations generated by these trajectory ensembles are crucial for understanding and quantizing many physical contexts, e.g., sheared fluids~\cite{Evans2004,Evans2010,Evans2004a}, dynamic phase transitions in glasses~\cite{Garrahan2007,Garrahan2009} and open quantum systems~\cite{Garrahan2010,Garrahan2011,Carollo2018}, and fluctuation theorems in nonequilibrium processes~\cite{Lebowitz1999,Kurchan2000,Maes1999,Gallavotti1995,Jarzynski1997,Crooks2000,Seifert2005}.

Generally, microcanonical trajectory ensembles can be regarded as a set of trajectories
in which time-integrated observables defined on trajectories are constrained by given values.
In contrast, in a canonical trajectory ensemble, constraints are imposed on the trajectory ensemble-averaged values of observables. Although these two ensembles seem to be very distinct, at long time limits and under certain conditions, they have been rigorously proven to be equivalent by using large deviation theory~\cite{chetrite2013,Chetrite2014,Touchette2008}. Importantly, auxiliary stochastic systems, including continuous diffusion~\cite{chetrite2013,Chetrite2014} and discrete jump stochastic systems~\cite{Jack2010,Garrahan2009,Garrahan2007}, have been explicitly constructed whose trajectories consist of the trajectories of canonical or microcanonical trajectory ensembles.

Recently, Carollo et al~\cite{Carollo2018} extended the results of previous research on classical stochastic systems to open quantum systems. The dynamics of open quantum systems obeys the time-independent quantum master equations. It has been well established that the quantum master equations can be unraveled into quantum trajectories and, in particular, that these trajectories have a classic probability interpretation~\cite{Breuer2002,Carmichael1993,Wiseman2010,Liu2018}. Therefore, this extension is not surprising. Inspired by their work, in this paper, we investigate the construction of an auxiliary quantum system for a periodically driven open quantum system~\cite{Kosloff2013,Szczygielski2013,Gasparinetti2014,Cuetara2015,GelbwaserKlimovsky2015,Liu2020}. Different from previous stationary equations, the dynamics of these open quantum systems is governed by the Floquet quantum master equation~\cite{Bluemel1991,Kohler1997,Breuer1997,Alicki2006}. Our physical situation of interest is very distinct from that of Carollo et al.~\cite{Carollo2018}; moreover, our theory is directly based on the notion of quantum trajectories and thus explicitly uses the probability formulas of quantum trajectories. Compared with the previous method that employs abstract operator calculations, our method is more similar to that of Chetrite and Touchette on classical stochastic systems~\cite{Chetrite2014}.

The remainder of this paper is organized as follows. In Sec.~(\ref{section2}), we review the Floquet quantum master equation and the unraveling of its quantum trajectories; the essential notations are defined therein. In Sec.~(\ref{section3}), we define the canonical quantum trajectory ensemble. In Sec.~(\ref{section4}), an auxiliary quantum system that can generate a canonical quantum trajectory ensemble is constructed for the finite time case and long time limit case. In Sec.~(\ref{section5}), a two-level quantum system is used to illustrate the procedure of constructing the auxiliary system in the long time limit case. Section~(\ref{section6}) concludes the paper.

\section{Floquet quantum master equation and trajectory unraveling}
\label{section2}
Assume that a periodically modulated quantum system interacts with a heat bath and that the inverse temperature of the heat bath is $\beta$. The Hamiltonian of the quantum system is $H(t)$ and $H\left(t+2\pi/\Omega\right)=H(t)$, where $\Omega$ is the driving frequency. According to the Floquet theorem~\cite{Zeldovich1967,Shirley1965}, the Hamiltonian satisfies an eigenvalue equation:
\begin{eqnarray}
\label{Floqueteigenvalueequation}
{\cal H}(t)|u_n(t)\rangle =\epsilon_n|u_n(t)\rangle,
\end{eqnarray}
where ${\cal H}(t)=H(t)-i\hbar\partial_t$ is the Floquet Hamiltonian and $\epsilon_n$ and $|u_n(t)\rangle$ ($n$$=$$1,\cdots,N$) are the quasi-energy and Floquet bases, respectively. Because the quasi-energy $\epsilon_n-q\hbar\Omega$ with basis $\exp(-iq\Omega t)|u_n(t)\rangle$ is also the solution of Eq.~(\ref{Floqueteigenvalueequation}), where $q$ is an arbitrary integer, we restrict the quasi-energies in a zone of size $\hbar\Omega$. Under the assumption of a weak system-bath coupling condition and appropriate time-scale separation, the evolution of the reduced density matrix of the quantum system $\rho(t)$ can be described by the Floquet quantum master equation~\cite{Grifoni1998,Breuer1997,Alicki2006}:
\begin{eqnarray}
\label{FQME}
\partial_{t}\rho(t)={\cal L}(t)[\rho(t)]=-\frac{i}{\hbar}[H(t),\rho(t)]+D(t)[\rho(t)].
\end{eqnarray}
The $D(t)$ term in the generator ${\cal L}(t)$ represents dissipation and dephasing due to the interaction between the system and the heat bath and is expressed as
\begin{eqnarray}
\label{dissipator}
D(t)[\rho]\nonumber =\sum_{\omega} r(\omega )\left[A(\omega ,t)\rho A^\dag(\omega ,t)- \frac{1}{2}
\left\{A^\dag (\omega ,t)A(\omega ,t),\rho\right\}\right],
\end{eqnarray}
where the summation is performed with respect to all possible Bohr frequencies $\omega$, which equal $(\epsilon_n-\epsilon_m)/\hbar + q\Omega$, and $q$ is a certain integer. The Bohr frequencies may be positive or negative but always appear in pairs. In the same equation, $A(\omega,t)$ and $A^\dag(\omega,t)$ are the Lindblad operators and are related by $
A^\dag(\omega,t)=A(-\omega,t)$ due to the Hermitian characteristics of the interaction Hamiltonian. Note that the Lindblad operators are also the eigenoperators of the Floquet Hamiltonian~\cite{Breuer1997}:
\begin{eqnarray}
\label{eigenoperatorFloquetHamiltonian}
[{\cal H}(t), A(\omega,t)]=-\hbar\omega A(\omega,t), \hspace{0.2cm} [{\cal H}(t), A^\dag(\omega,t)]=\hbar\omega A^\dag(\omega,t).
\end{eqnarray}
Because the heat bath is always in the thermal state with the inverse temperature $\beta$, the Fourier transformation $r(\omega)$ of the correlation function of the heat bath satisfies the Kubo-Martin-Schwinger (KMS) condition~\cite{Breuer2000}: $r(-\omega )=r(\omega )\exp(-\beta \hbar\omega)$.

Eq.~(\ref{FQME}) can be unraveled into the dynamics of individual quantum systems~\cite{Carmichael1993,Plenio1998,Breuer2000,Wiseman2010}. The evolution of each system is alternately composed of continuous processes and discrete random jumps. Assume that the jumps occur at time $t_i$ with Bohr frequency $\omega_i$, where $i=1,\cdots,M$ and $M$ denotes the total number of jumps. When the evolution ends at time $T$, a quantum trajectory is generated and is denoted as $\overrightarrow \omega_M=\{\omega_1,\cdots,\omega_M\}$. Further assuming that the density matrixes of these individual quantum systems at time $t$ ($\le T$) are $\widetilde\rho(\overrightarrow \omega_M,t)$, the solution of Eq.~(\ref{FQME}) is equal to a quantum trajectory ensemble average:
\begin{eqnarray}
\label{probabilityformofdensitymatrix}
\rho(t)
&=&\sumint_{M=0}^\infty {\cal D}(t)\hspace{0.1cm}  p(\overrightarrow{\omega}_M,t ) {\widetilde \rho}(\overrightarrow{\omega}_M,t),
\end{eqnarray}
where the sum-integral symbol means that we sum with respect to all possible quantum trajectories and integrate over all possible times~\cite{Liu2016a,Liu2018,Liu2020},
\begin{eqnarray}
\label{densitymatrixquantumjumptraj}
{\widetilde \rho}(\overrightarrow{\omega}_M,t)=\frac{1 }{p(\overrightarrow{\omega}_M ,t)}
G_{t,t_M}\left( J_{t_M}\left(G_{t_M,t_{M-1}}\left(\cdots J_{t_1} \left( G_{t_1,0}\left(\rho_0\right)\right)\cdots\right)\right)\right),
\end{eqnarray}
where $\rho_0$ is the density matrix at initial time $0$ and the denominator $p(\overrightarrow{\omega}_M,t)$ is the probability distribution of observing the quantum trajectory $\overrightarrow{\omega}_M$, which is simply equal to the trace of the numerator.
In Eq.~(\ref{densitymatrixquantumjumptraj}), the symbols $G$ and $J$ denote superoperators, and their action regions are explicitly indicated by the round brackets. The superoperator $G_{t_i,t_{i-1}}={\cal T}_{-}\exp[\int_{t_{i-1}}^{t_i}d\tau{L}_0(\tau)]$ and its generator is
\begin{eqnarray}
\label{continuouspart}
 L_{0}(t)[O]&=&-\frac{i}{\hbar}[H(t),O]-\frac{1}{2}  \sum r(\omega )\left\{A^\dag(\omega,t)A(\omega,t),O\right\},
\end{eqnarray}
where ${\cal T}_-$ is the time-ordering operator. $J_{t_i}(O)$ is a shorthand expression of the jump superoperator 
$r(\omega_i)A(\omega_i,t_i)O A^\dag(\omega_i,t_i)$.

The quantum trajectory ensemble that we are interested in is as follows. At initial time $0$, the individual quantum systems are in one of the Floquet pure states $|u_m(0)\rangle \langle u_m(0)|$, which is randomly selected based on a probability distribution $p_m$, $m=1,\cdots,N$. That is, the quantum ensemble is initially in a mixed state: $\rho_0=\sum_{m=1}^N p_m |u_m(0)\rangle \langle u_m(0)|$. The quantum system evolves, and a quantum trajectory $\overrightarrow{\omega}_M$ is recorded. At the end time $T$, the system is measured in the Floquet basis, and a pure state $|u_n(T)\rangle\langle u_n(T)|$ is obtained. It is not difficult to argue that, if the joint probability distribution of observing the quantum trajectory with the special initial and terminal bases is $p_{m,n}(\overrightarrow{\omega}_M,T)=p_mp_{m|n}(\overrightarrow{\omega}_M,T)$, the conditional probability distribution equals~\cite{Liu2016a,Liu2018}
\begin{eqnarray}
\label{probdensityfunctional}
p_{m|n}(\overrightarrow{\omega}_M,T)=\langle u_n(T)|G_{T,t_M}\left( J_{t_M}\left(G_{t_M,t_{M-1}}\left(\cdots J_{t_1} \left( G_{t_1,0}\left( |u_m(0) \rangle \langle u_m(0)| \right)\right)\cdots\right)\right)\right) |u_n(T)\rangle.
\end{eqnarray}

\section{Canonical quantum trajectory ensemble}
\label{section3}
In this paper, we investigate canonical quantum trajectory ensembles~\cite{Garrahan2009,chetrite2013,Chetrite2014}. To define such an ensemble, we choose a time-integrated observable of interest, namely, the stochastic heat current along quantum trajectories. According to the interpretation of quantum measurements~\cite{Breuer1997,Breuer2002,Liu2018}, the occurrence of jump events along a quantum trajectory $\overrightarrow \omega_M=\{\omega_1,\cdots,\omega_M\}$ indicates that the quantum system exchanges quanta $\hbar\omega_i$ with the heat bath. From a thermodynamic point of view, these quanta represent the discrete heat released to the environment~\cite{Breuer2003,DeRoeck2006,Crooks2008,Horowitz2012,Hekking2013,Manzano2015,Liu2016a}. Hence, given a quantum trajectory $\overrightarrow \omega_M$ with duration $T$, the heat current is equal to $j=Q_T/T$, and the total heat production is
\begin{eqnarray}
\label{stocheat}
Q_T(\overrightarrow{\omega}_M)=\sum_{i=1}^M \hbar\omega_i.
\end{eqnarray}
Note that our object of interest is different from the average rate of quantum jumps, which was the focus of Carollo et al.~\cite{Carollo2018}. Nevertheless, the extension of our formulas to the latter case is direct. Given the observable, in a canonical quantum trajectory ensemble, the probability distribution of the quantum trajectory $\overrightarrow{\omega}_M$ with the initial Floquet basis $|u_m(0)\rangle$ and terminal Floquet basis $|u_n(T)\rangle$ is~\cite{Garrahan2009,chetrite2013,Chetrite2014}
\begin{eqnarray}
\label{canonicaldistribution}
p_{m,n}({\chi},\overrightarrow{\omega}_M,T)=\frac{1}{\Phi_T(\chi)}p_{m,n}(\overrightarrow{\omega}_M,T)e^{\chi Q_T(\overrightarrow{\omega}_M)},
\end{eqnarray}
where the normalized factor $\Phi_T(\chi)$ is the moment generating function,
\begin{eqnarray}
\label{heatmomentgenreatingfunction}
\Phi_T (\chi)&=&\sum_{m,n=1}^N\sumint_{M=0}^{\infty} {\cal D}(t)\hspace{0.1cm}  p_{m,n}(\overrightarrow{\omega}_M,T) e^{\chi Q_T(\overrightarrow{\omega})}. 
\end{eqnarray}

\section{Auxiliary open quantum system}
\label{section4}
\subsection{Finite time case}
Now, we are in a position to construct an auxiliary open quantum system whose quantum trajectory ensemble is consistent with Eq.~(\ref{canonicaldistribution}).
Assume the Hamiltonian and Lindblad operators of the auxiliary system are $H'(t)$ and $A'(\omega,t)$, respectively. In the remainder of this paper, we denote quantities in the auxiliary system with a prime symbol unless otherwise stated. Inspired by Carollo et al.~\cite{Carollo2018}, we set
\begin{eqnarray}
\label{LindbladAuxiliarysystem}
A'(\omega,t)=l_\chi(t)A(\omega,t)l_\chi^{-1}(t) e^{\chi\hbar\omega/2},
\end{eqnarray}
where $l_\chi(t)$ is an invertible Hermitian operator to be solved~\footnote{This result is easy to obtain if one considers the consistency between the classical jump master equation and the diagonal part of the quantum master equation~(\ref{FQME}) represented in the Floquet basis~\cite{Garrahan2009,Jack2010}. }. We preliminarily require that the operator depends on the parameter $\chi$. Obviously, Eq.~(\ref{LindbladAuxiliarysystem}) implies that the jump superoperator of the auxiliary system is related to that of the original system as follows:
\begin{eqnarray}
\label{jumppartauxiliarysystem}
J'_{t_i}(O)=e^{\chi\hbar\omega_i}l_{t_i}J_{t_i}(l^{-1}_{t_i}Ol^{-1}_{t_i})l_{t_i}.
\end{eqnarray}
For simplicity, we denote $l_{t_i}=l_\chi(t_i)$ and $l_{t_i}^{-1}=l_\chi^{-1}(t_i)$. Substituting Eq.~(\ref{jumppartauxiliarysystem}) into $p'_{m,n}(\overrightarrow{\omega}_M,T)$, we obtain the probability distribution of observing the quantum trajectory $\overrightarrow{\omega}_M$ with the initial basis $|u_m(0)\rangle$ and terminal basis $|u_n(T)\rangle$ in the auxiliary open quantum system:
\begin{eqnarray}
p'_{m,n}(\overrightarrow{\omega}_M,T)&=&e^{\chi Q(\overrightarrow{\omega}_M)} \langle u_n(T)| l_T l^{-1}_T
G'_{T,t_M}\left(l_{t_M} J_{t_M} \left( l_{t_M}^{-1} G'_{t_M,t_{M-1}}\left(l_{t_{M-1}} \right.\right.\right. \nonumber \\
&&\left. \left.\left. \cdots J_{t_1}\left(l_{t_1}^{-1} G'_{t_1,0}\left(l_0 l^{-1}_0|u_m(0) \rangle \langle u_m(0)|  l^{-1}_{0}l_0 \right)l_{t_1}^{-1}\right) \cdots  l_{t_{M-1}}\right) l_{t_M}^{-1}\right)l_{t_M}\right)l^{-1}_Tl_T |u_n(T)\rangle p_m.
\end{eqnarray}
Compared with Eq.~(\ref{canonicaldistribution}), we find that, if
\begin{eqnarray}
\label{continouspartauxiliarysystem}
&&l^{-1}_{t_i}G'_{t_i,t_{i-1}}(l_{t_{i-1}}O l_{t_{i-1}})l^{-1}_{t_i}=e^{-\Lambda_T(\chi)(t_i-t_{i-1})} G_{t_i,t_{i-1}}(O), \\
&&\hspace{0.2cm} l_{\chi}(0)|u_m(0)\rangle =|u_m(0)\rangle, \hspace{0.2cm} {\rm and} \hspace{0.2cm} l_{\chi}(T)|u_n(T)\rangle=|u_n(T)\rangle,
\label{initialandendconditions}
\end{eqnarray}
where
\begin{eqnarray}
\Lambda_T(\chi)=\frac{1}{T}\ln\Phi_T(\chi)
\end{eqnarray}
is the scaled cumulant generating function~\cite{Touchette2008}, the auxiliary system will generate the same quantum trajectory ensemble as the canonical one. Simple derivations further show that Eq.~(\ref{continouspartauxiliarysystem}) is also equivalent to the following conditions:
\begin{eqnarray}
\label{Hamiltonianauxiliarysystem}
H'(t)&=&\frac{1}{2}l_\chi(t)H(t)l_\chi^{-1}(t)  +\frac{i\hbar}{2}\partial_t l_\chi(t) l_\chi^{-1}(t) +\frac{i\hbar}{4}l_\chi^{-1}(t)\left[
\sum_\omega r(\omega) A^\dag(\omega,t)A(\omega,t)\right]l_\chi(t) + H.c.,
\end{eqnarray}
where $H.c.$ denotes Hermitian conjugation and $l_\chi(t)$ must satisfy the operator equation given by
\begin{eqnarray}
\label{leftvector}
\partial_t  l_\chi^2(t)+ {\cal L}_\chi^*(t)[l_\chi^2(t)]=\Lambda_T(\chi) l_\chi^2(t),
\end{eqnarray}
where
\begin{eqnarray}
{\cal L}_\chi^*(t)(O)=\frac{i}{\hbar}[H(t), O]+\sum_{\omega} r(\omega )\left[e^{\chi\hbar\omega}A^\dag(\omega ,t)O A(\omega ,t)- \frac{1}{2}
\left\{A^\dag (\omega ,t)A(\omega ,t),O\right\}\right].
\end{eqnarray}
The $\chi$-dependence of the operator $l_\chi(t)$ is explicit. Using Eqs.~(\ref{Hamiltonianauxiliarysystem}) and~(\ref{LindbladAuxiliarysystem}), we can rewrite the generator of the quantum master equation of the auxiliary quantum system as
\begin{eqnarray}
\label{generatorauxiliarysystem}
{\cal L}'(t)(O)&=&-\frac{i}{\hbar}[H'(t),O]+\sum_{\omega} r(\omega )\left[A'(\omega ,t)O A'^\dag(\omega ,t)- \frac{1}{2}
\left\{A'^\dag (\omega ,t)A'(\omega ,t),O\right\}\right] \nonumber \\
&=&l_\chi(t){\cal L}_\chi(t)\left[ l_\chi^{-1}(t)Ol_\chi^{-1}(t) \right]l_\chi(t)-Ol_\chi^{-1}(t) {\cal L}^*_\chi(t)\left[ l_\chi^{2}(t) \right] l_\chi^{-1}(t) - O (\partial_t l_\chi(t)) l_\chi^{-1}(t) +(\partial_t l_\chi(t))l_\chi^{-1}(t)O\nonumber \\
&=&l_\chi(t) {\cal L}_\chi(t)\left[ l_\chi^{-1}(t)Ol_\chi^{-1}(t) \right] l_\chi(t)-\Lambda_T(\chi)O + Ol_\chi^{-1}(t) (\partial_t l_\chi(t))+(\partial_t l_\chi(t))l_\chi^{-1}(t)O,
\end{eqnarray}
where ${\cal L}_\chi(t)$ is the dual of ${\cal L}^*_\chi(t)$ and is defined as ${\rm Tr}[O_1{\cal L}_\chi(t)(O_2)]={\rm Tr}[{\cal L}^*_\chi(t)(O_1)O_2]$, and the third equation is a consequence of applying Eq.~(\ref{leftvector}). This is a time-dependent quantum Doob transform. We note that
the generator of the quantum master equation
is slightly distinct from that obtained by Carollo et al.~\cite{Carollo2018}, in which the second term about the scaled generating function $\Lambda_T(\chi)$ is absent.


\subsection{Long time limit case}
Although the previous arguments are generally valid and the equations are formally correct, it is not very clear whether Eq.~(\ref{leftvector}) always achieves a solution that can satisfy both the initial and the terminal conditions of Eq.~(\ref{initialandendconditions}). In addition, these results do not depend on any special properties of Floquet open quantum systems. Now we focus on the long time limit case. An interesting feature of these open quantum systems is that, at a long time limit, Eq.~(\ref{FQME}) evolves into a periodic limit cycle $\rho(T)=\rho(T+2\pi/\Omega)$~\cite{Breuer2000,Szczygielski2013,Grifoni1998} and
\begin{eqnarray}
\label{longtimedensitymatrix}
\lim_{T\rightarrow\infty} \rho(T)=\sum_{n=1}^N P_n|u_n(T)\rangle\langle u_n(T)|,
\end{eqnarray}
where $P_n$, $n=1,\cdots,N$, is the stationary probability distribution, which is uniquely determined by the generator of Eq.~(\ref{FQME}). Importantly, under the same situation, the stochastic heat~(\ref{stocheat}) obeys the large deviation principle~\cite{Touchette2008,Gasparinetti2014,Cuetara2015,Liu2020}: $p(j)\asymp\exp[-TI(j)]$, where the symbol $\asymp$ denotes an asymptotically exponential approximation and $I(j)$ is the rate function. The scaled cumulant generating function is
\begin{eqnarray}
\label{scaledcumulantgeneratingfunction}
\lim_{T\rightarrow\infty} \Lambda_T(\chi)=\Lambda(\chi),
\end{eqnarray}
where the time-independent $\Lambda(\chi)$ is the maximum eigenvalue of the generator ${\cal L}_\chi(t)$.
The rate function and scaled cumulant generating function are related
by the Legendre transform,
\begin{eqnarray}
\label{Legendre}
I(j)=\max_{\chi}\{j\chi- \Lambda(\chi)\}.
\end{eqnarray}
Appendix A explains Eq.~(\ref{longtimedensitymatrix}) and the large deviation principle.
Hence, we introduce the logarithmic equivalence between the canonical quantum trajectory ensemble and the trajectory ensemble generated by the auxiliary quantum system as that in classical stochastic systems~\cite{Garrahan2009,Jack2010,chetrite2013,Chetrite2014}:
\begin{eqnarray}
\label{logarithmequivalence}
\lim_{T\rightarrow\infty} \frac{1}{T}\ln\frac{
p'_{m,n}(\overrightarrow{\omega}_M,T)}{
p_{m,n}({\chi},\overrightarrow{\omega}_M,T) }\rightarrow 0.
\end{eqnarray}
Correspondingly, the conditions~(\ref{initialandendconditions}) are relaxed to
\begin{eqnarray}
\label{relaxedinitialandendconditions}
l_\chi(0)|u_m(0)\rangle \propto |u_m(0)\rangle \hspace{0.2cm}{\rm and}\hspace{0.2cm} l_\chi(T)|u_n(T)\rangle \propto |u_n(T)\rangle.
\end{eqnarray}
Then, the relationship between the moment generating functions of the auxiliary and original quantum systems can be intuitively obtained as
\begin{eqnarray}
{\Phi}'_T(\chi')&=&\sum_{m,n=1}^N \sumint_{M=0}^\infty {\cal D}t \hspace{0.1cm }e^{\chi' Q_T(\overrightarrow{\omega})}p'_{m,n}(\overrightarrow{\omega}_M,T) \nonumber \\
&=&\frac{1}{\Phi_T(\chi)}\sum_{m,n=1}^N \sumint_{M=0}^\infty {\cal D}t \hspace{0.1cm }e^{(\chi'+\chi) Q_T(\overrightarrow{\omega})}\frac{p'_{m,n}(\overrightarrow{\omega}_M,T)}{  p_{m,n}(\chi,\overrightarrow{\omega}_M,T) } p_{m,n}(\overrightarrow{\omega}_M,T).
\end{eqnarray}
When the duration $T$ tends to infinity and we let the scaled cumulant generating function of the auxiliary quantum system be $\Lambda'(\chi')$, because of the logarithmic equivalence~(\ref{logarithmequivalence}), we immediately have
\begin{eqnarray}
\label{SCGFauxiliarysystem}
{\Lambda}' (\chi')=\Lambda(\chi'+\chi)-\Lambda(\chi).
\end{eqnarray}
This agrees with the crucial result obtained by Carolla et al. for the time-independent quantum master equations~\cite{Carollo2018} (see Eq.~(10) therein). Due to the Legendre transform~(\ref{Legendre}) and Eq.~(\ref{SCGFauxiliarysystem}), the typical heat current of the auxiliary quantum system $\partial_{\chi'}\Lambda'|_{\chi'=0}$, which is equal to the average heat current $\langle j'\rangle$, is equal to $\partial_{\chi}\Lambda(\chi)$. Note that the latter is also the average heat current of the canonical quantum trajectory ensemble. Because of the equivalence of the canonical and microcanonical trajectory ensembles, if $\chi\neq 0$,
the typical heat current of the auxiliary quantum system becomes
the atypical heat current of the original quantum system conditioned on this large deviation~\cite{Garrahan2009,Jack2010,chetrite2013}.

Eqs.~(\ref{logarithmequivalence})-(\ref{SCGFauxiliarysystem}) depend on the operator $l_\chi(t)$, which satisfies both Eqs.~(\ref{leftvector}) and~(\ref{relaxedinitialandendconditions}). Here, we argue
the existence of this operator.
We conjecture that the solution of Eq.~(\ref{leftvector}) is diagonal in the Floquet basis, and its elements are time-independent:
\begin{eqnarray}
\label{l2expression}
l_\chi^2(t)=\sum_{n=1}^N l^2_n(\chi)|u_n(t)\rangle \langle u_n(t)|.
\end{eqnarray}
Apparently, this solution satisfies Eq.~(\ref{relaxedinitialandendconditions}). In addition, the operator is periodic. Substituting the solution into Eq.~(\ref{leftvector}) and writing the equation in the Floquet basis, we obtain a matrix equation:
\begin{eqnarray}
\label{lefteigenvector}
\widetilde{\bf R}(\chi) \overrightarrow {l^2}(\chi) =\Lambda(\chi)  \overrightarrow{l^2}(\chi),
\end{eqnarray}
where the column vector $\overrightarrow{ l^2}(\chi)$ is the transpose of the row vector $(l^2_1(\chi),\cdots,l^2_N(\chi))$ and the tilde symbol represents the transpose; the definition of the matrix ${\bf R}(\chi)$ is presented in Appendix A.
Eq.~(\ref{lefteigenvector}) is simply the left eigenvector $\overrightarrow{ l^2}(\chi)$ of the matrix ${\bf R}(\chi)$ with the maximum eigenvalue $\Lambda(\chi)$. Their positivity is ensured by the definition of the moment generating function and the Perron-Frobenius theorem. Appendix A gives some details.

The diagonal structure of $l_\chi^2(t)$ leads to an interesting consequence. First, the formally complicated Hamiltonian $H'(t)$ of the auxiliary quantum system is simply equal to the Hamiltonian $H(t)$ of the original quantum system. The proof includes two steps. In the first step, Eq.~(\ref{l2expression}) implies that the sum of the first two terms and their Hermitian conjugations in Eq.~(\ref{Hamiltonianauxiliarysystem}) is
\begin{eqnarray}
&&\frac{1}{2}l_\chi(t)H(t)l_\chi^{-1}(t)  +\frac{i\hbar}{2}\partial_t l_\chi(t) l_\chi^{-1}(t)+H.c\nonumber\\
&=&\frac{1}{2}\sum_{n=1}^N ( i\hbar \partial_t |u_n(t)\rangle + \epsilon_n|u_n(t)\rangle) \langle u_n(t)|+H.c\nonumber\\
&=&H(t).
\end{eqnarray}
For this purpose, we use Eq.~(\ref{Floqueteigenvalueequation}) and the completeness of the Floquet basis. The second step makes use of the microscopic expressions of the Lindblad operators. Assume the interaction Hamiltonian between the quantum system and the heat bath to be $A\otimes B$, where $A$ and $B$ are the Hermitian operators of the system and heat bath, respectively. Then, the Lindblad operator is~\cite{Breuer1997}
\begin{eqnarray}
\label{Amatrix}
A(\omega,t)&=&\sum_{m,n,q}\delta_{\omega,\epsilon_n-\epsilon_m + q\Omega}\langle\langle u_m|A|u_n\rangle \rangle_q |u_m(t)\rangle\langle u_n(t)|e^{-iq\Omega t},
\end{eqnarray}
where $\delta$ is the Kronecker symbol and the time-independent coefficient $\langle\langle u_m|A|u_n\rangle \rangle_q $ is the $q$-th harmonic of the transition amplitude $\langle u_m(t)|A|u_n(t) \rangle$, i.e.,
\begin{eqnarray}
\langle\langle u_m|A|u_n\rangle \rangle_q =\frac{\Omega}{2\pi}\int_0^{2\pi/\Omega} \langle u_m(t)|A|u_n(t) \rangle e^{iq\Omega t}.
\end{eqnarray}
Using this expression and performing a straightforward argument, we find that the square bracket term in Eq.~(\ref{Hamiltonianauxiliarysystem}) is diagonal and especially real in the Floquet basis. Because the operator $l_\chi(t)$ is also diagonal, we conclude that the sum of the third term and its Hermitian conjugation is exactly zero.
In addition, we can also verify that the Lindblad operators $A'(\omega,t)$ and $A'^\dag(\omega,t)$ of the auxiliary quantum system are still the eigenoperators of the Floquet Hamiltonian ${\cal H}(t)$.

Before closing this section, we want to point out that the scaled cumulant generating function $\Lambda'(\chi')$ of the auxiliary quantum system satisfies a matrix equation analogous to Eq.~(\ref{lefteigenvector}):
\begin{eqnarray}
\label{lefteigenvectorauxiliarysystem}
\widetilde{\bf R'}(\chi') \overrightarrow {l'^2}(\chi') =\Lambda'(\chi')  \overrightarrow{l'^2}(\chi'),
\end{eqnarray}
where the matrix $\widetilde{\bf R'}(\chi')$ is determined by the generator ${\cal L}'_{\chi'}(t)$. Although we have obtained Eq.~(\ref{SCGFauxiliarysystem}), it shall be interesting to see how the same conclusion is achieved and what the relation is between $ \overrightarrow {l'^2}(\chi')$ and $ \overrightarrow {l^2}(\chi)$ from an eigenvalue matrix equation perspective. Some discussion of this topic is presented in Appendix B.

\section{Two-level quantum system }
\label{section5}
In this section, we use a two-level quantum system driven by a periodic external field~\cite{Breuer1997,Szczygielski2013,Langemeyer2014,Gasparinetti2014,Cuetara2015,Liu2020} to illustrate the construction of the auxiliary quantum system at a long time limit. The Hamiltonian is given as
\begin{eqnarray}
\label{TLS}
H(t)=\frac{\hbar{\omega_0}}{2} \sigma_z +\frac{\hbar\Omega_R}{2}\left(\sigma_+ e^{-i\Omega t}+\sigma_- e^{i\Omega t}\right),
\end{eqnarray}
where $\omega_0$ is the transition frequency of the bare system, $\Omega_R$ is the Rabi frequency, and $\Omega$ is the frequency of the external field. The Floquet basis and quasi-energy are
\begin{eqnarray}
\label{TLSFloquetbases}
|u_{\pm}(t)\rangle =\frac{1}{\sqrt{2\Omega'}}
\left(\begin{array}{c}
 \pm \sqrt{\Omega'\pm\delta}\\
 e^{i\Omega t}\sqrt{\Omega'\mp\delta},
\end{array}\right),
\end{eqnarray}
and $\epsilon_\pm=\hbar(\Omega \pm \Omega')/2$, respectively, where $\Omega'=\sqrt{\delta ^2+\Omega_R^2}$ and the detuning parameter $\delta=\omega_0-\Omega$. Here, we additionally set $\Omega>\Omega'$. 
We assume that the coupling between the quantum system and heat bath is $\sigma_x$-coupling. There are six Lindblad operators: three of them with Bohr frequencies $\Omega $, $(\Omega-\Omega')$, and $(\Omega+\Omega')$ are
\begin{eqnarray}
\label{Lindbladoperatorsorg1}
A(\Omega,t)&=&\frac{\Omega_R}{2\Omega'}\left(|u_+(t)\rangle\langle u_+(t)|-|u_-(t)\rangle\langle u_-(t)| \right)e^{-i\Omega t},\nonumber\\
A(\Omega -\Omega',t)&=&\left(\frac{\delta-\Omega'}{2\Omega'} \right)|u_+(t)\rangle\langle u_-(t)|e^{-i\Omega t},\\
A(\Omega+\Omega',t)&=&\left(\frac{\delta+\Omega'}{2\Omega'} \right) |u_-(t)\rangle\langle u_+(t)| e^{-i\Omega t};\nonumber
\end{eqnarray}
the other three Lindblad operators with Bohr frequencies $-\Omega $, $-(\Omega-\Omega')$, and $-(\Omega+\Omega')$ are the adjoint operators of Eq.~(\ref{Lindbladoperatorsorg1}). Through a simple derivation, we obtain the matrix elements of ${\textbf R}(\chi)$:
\begin{eqnarray}
\label{twolevelmatrixes}
{R}_{11}(\chi)&=& (e^{\chi\hbar\Omega}-1)\Gamma_{+\Omega}+(e^{-\chi\hbar\Omega}-1)\Gamma_{-\Omega}-\Gamma_{-(\Omega-\Omega')}-
\Gamma_{+(\Omega+\Omega')}, \nonumber\\
 {R}_{22}(\chi)&=&(e^{\chi\hbar\Omega}-1)\Gamma_{+\Omega}  + (e^{-\chi\hbar\Omega}-1)\Gamma_{ -\Omega} -\Gamma_{+(\Omega -\Omega')}-
\Gamma_{-(\Omega+\Omega')},\nonumber \\
  {R}_{12}(\chi)&=&e^{ \chi\hbar(\Omega-\Omega')}\Gamma_{+(\Omega-\Omega')} + e^{- \chi\hbar(\Omega+\Omega')}\Gamma_{ -(\Omega+\Omega')}, \nonumber \\
 {R}_{21}(\chi)&=&e^{-\chi\hbar(\Omega-\Omega')}\Gamma_{-(\Omega-\Omega')}+ e^{\chi\hbar(\Omega+\Omega')}\Gamma_{ +(\Omega+\Omega')},
\end{eqnarray}
where the coefficients are
\begin{eqnarray}
\label{coeffs}
\Gamma_{\pm\Omega}&=&\left(\frac{\Omega_R}{2\Omega'}\right)^2 r(\pm\Omega), \nonumber \\
\Gamma_{\pm(\Omega-\Omega')} &=&\left(\frac{\delta-\Omega'}{2\Omega'}\right)^2 r(\pm(\Omega-\Omega')), \\
\Gamma_{\pm(\Omega+\Omega')}&=&\left(\frac{\delta+\Omega'}{2\Omega'}\right)^2 r(\pm(\Omega+\Omega')).\nonumber
\end{eqnarray}
Because the above is a simple $2\times 2$ matrix, we can easily write its maximum eigenvalue and the corresponding left eigenvector: $\Lambda(\chi)=[{R}_{11}(\chi)+{R}_{22}(\chi)+B(\chi)]/2$ and
\begin{eqnarray}
(l^2_1(\chi),l^2_2(\chi))=\left(\frac{{R}_{11}(\chi)- {R}_{22}(\chi)+B(\chi)}{2{R}_{12}(\chi)},1\right),
\end{eqnarray}
where $B(\chi)=\sqrt{[ {R}_{11}(\chi)- {R}_{22}(\chi)]^2+4 {R}_{21}(\chi){R}_{12}(\chi)}$.
Based on these results, we can directly construct the auxiliary open quantum system. For instance, three of the Lindblad operators are
\begin{eqnarray}
\label{Lindbladoperatorsauxiliary1}
A'(\Omega,t)&=&e^{\chi\hbar\Omega/2}A(\Omega,t),\\
A'(\Omega -\Omega',t)&=&e^{\chi\hbar(\Omega-\Omega')/2}\frac{l_1(\chi)}{l_2(\chi)}A(\Omega -\Omega',t),\\
A'(\Omega+\Omega',t)&=&e^{\chi\hbar(\Omega+\Omega')/2}\frac{l_2(\chi)}{l_1(\chi)} A(\Omega+\Omega',t).
\label{Lindbladoperatorsauxiliary3}
\end{eqnarray}
We must emphasize that the other three Lindblad operators $A'(\omega,t)$ with $\omega=-\Omega $, $-(\Omega -\Omega')$, and $-(\Omega +\Omega')$ are equal to the adjoints of Eq.~(\ref{Lindbladoperatorsauxiliary1})-~(\ref{Lindbladoperatorsauxiliary3}) and that $\chi$ therein is replaced by $-\chi$. For the long time limit case, the Hamiltonian of the auxiliary quantum system is simply Eq.~(\ref{TLS}).

\begin{figure}
\includegraphics[width=1.\columnwidth]{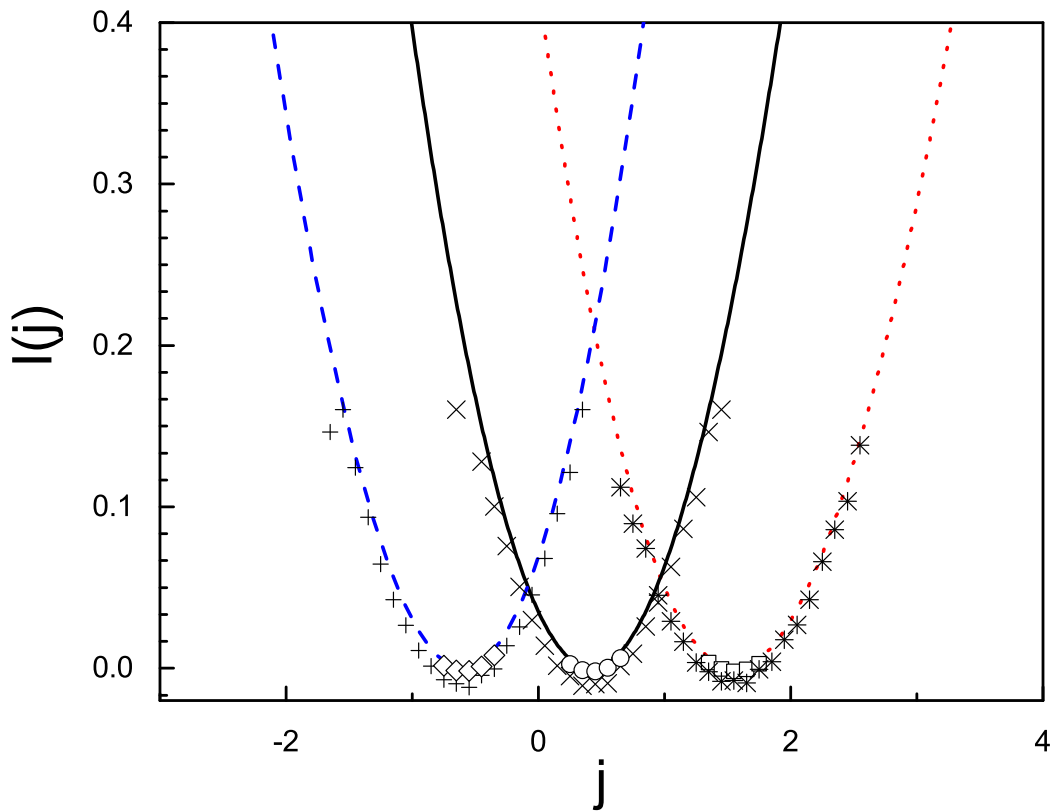}
\caption{Large deviation functions $I(j)$ of the original (solid curve) and auxiliary two-level quantum systems with $\chi=-0.4$ (dashed curve) and $0.4$ (dotted curve). The functions are solved by applying the Legendre transform~(\ref{Legendre}) to Eq.~(\ref{SCGFauxiliarysystem}). The symbols are the approximated large deviation functions and are calculated by $-\ln p(j)/T$, where the probability distribution $p(j)$ is collected by simulating the quantum trajectories. The durations for the cross and hollow symbols are 100 and 1000, respectively. The Fourier transforms of the correlation functions are set to be $r(\omega)$$=$${\cal A}|{\omega}|^3 {\cal N}_k(\omega)$ for $\omega$$<$$0$; otherwise, $r(\omega)$$=$${\cal A}|{\omega}|^3 [{\cal N}_k(\omega)+1]$, where ${\cal N}(\omega)$$=$$1/[\exp ( \beta |\omega| )-1]$ and the coefficient $A$ is related to the coupling strength between the system and the heat bath~\cite{Breuer2000}. The parameters used are $\omega_0$$=$$1$, $\Omega_R$$=$$0.8$, $\Omega$$=$$1.1$, ${\cal A}$$=$$1$, and $\beta=1/3$. }
\label{fig1}
\end{figure}

\begin{figure}
\includegraphics[width=1.\columnwidth]{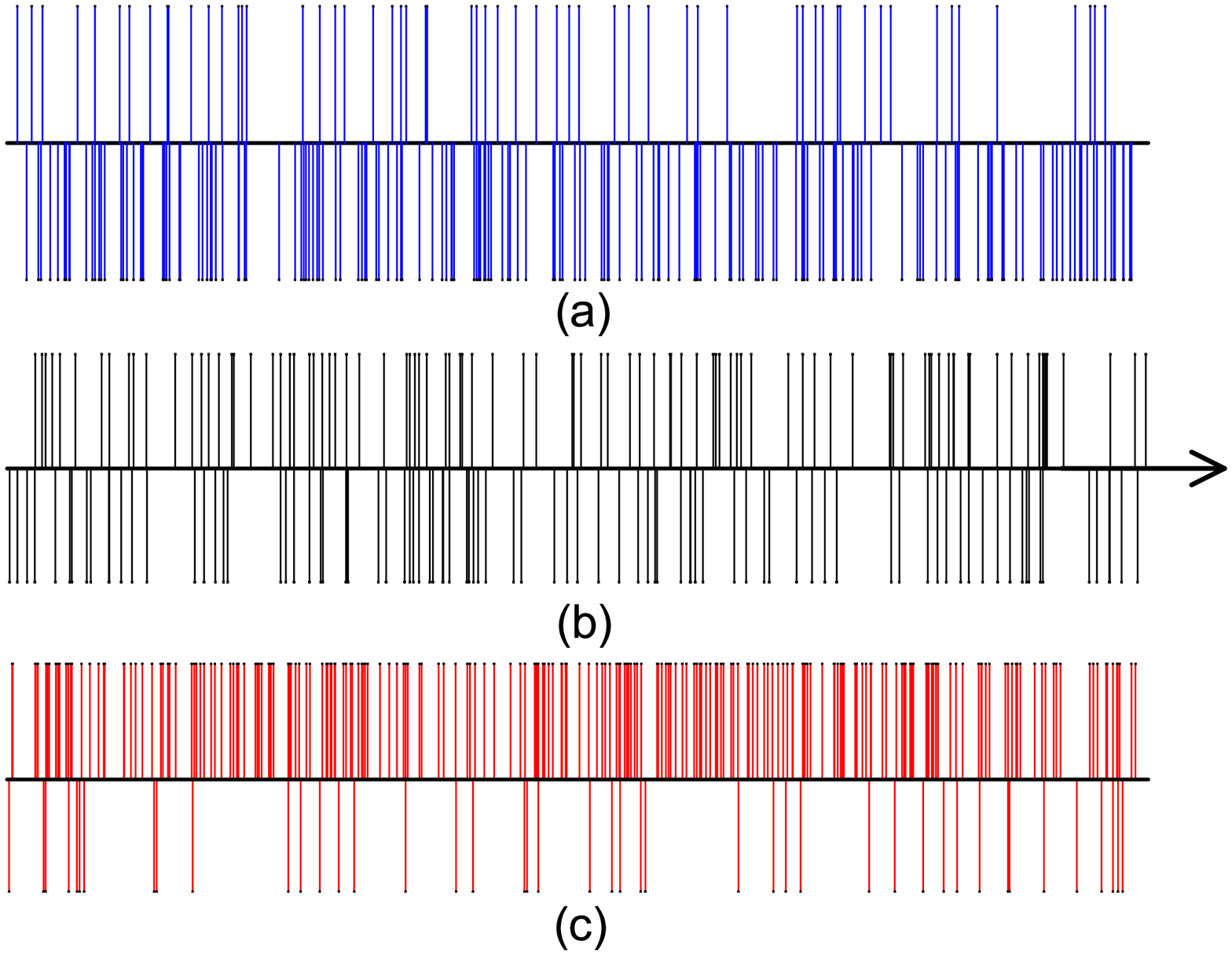}
\caption{Segments of the simulated quantum trajectories, where the $\chi$ values in panels (a), (b), and (c) are $1$, $0.$, and $-1$, respectively, and the time intervals are 50 after the system is in a steady state. The other parameters are the same as those used in Fig.~(\ref{fig1}). The arrow in the middle panel denotes the time orientation. The short lines denote time points at which discrete quantum jumps occur. The upper and bottom lines represent the released and absorbed heat, respectively. }
\label{fig2}
\end{figure}

Fig.~\ref{fig1} shows the rate functions of the original quantum system and auxiliary quantum system. These functions are solved by applying the Legendre transform~(\ref{Legendre}) to the scaled cumulant generating functions (the curves) and by simulating the quantum trajectories (the symbols). We can clearly see that, by modulating $\chi$, the typical current of the auxiliary system, i.e., the value at the minimum of its rate function, is located at different large deviations from the original system.
To intuitively express the behaviors of the auxiliary quantum system, we also plot several segments of the quantum trajectories from the original and auxiliary quantum systems in Fig.~\ref{fig2}.
Figure~(\ref{fig1}) reminds us that, by choosing a sufficiently negative $\chi$, the typical current can become negative. Then, along the quantum trajectories, one would observe that heat is more frequently absorbed from than released into the heat bath. Figure~(\ref{fig2})(a) confirms this expectation. When we examine the algorithm of simulating the quantum trajectories~\cite{Liu2016a}, we note that a negative $\chi$ value exponentially inhibits the rates of quantum jumps with heat release and exponentially increases the rates of quantum jumps with heat absorption (data not shown). These $\chi$-modulated rates do not satisfy the KMS condition. This observation simply explains the relation between panels (a) and (b). If $\chi$ is positive, the opposite will happen; see Fig.~(\ref{fig2})(c).

Let us comment on the typical negative heat current to finish this work. According to the first law of thermodynamics, a positive power is output from the heat bath. Therefore, the auxiliary quantum system seems to violate the second law of thermodynamics. However, previous studies have confirmed that Floquet open quantum systems strictly obey the law of thermodynamics~\cite{Szczygielski2013,Cuetara2015,Gasparinetti2014,Langemeyer2014,Liu2020}. In fact, there is no contradiction because the auxiliary quantum system is not physical in the sense that ${A'}^\dag(\omega,t)\neq A'(-\omega,t)$; see their definitions in Eq.~(\ref{LindbladAuxiliarysystem}).

\section{Conclusion}
\label{section6}
In this paper, we investigate the construction of an auxiliary open quantum system that can generate a canonical quantum trajectory ensemble. The dynamics of the original quantum system is governed by the Floquet quantum master equation. The probability formula of the quantum trajectory is explicitly used. We see that the theory on quantum trajectory ensembles is highly analogous to the theory on classical  trajectory ensembles. Since most of the studies in the literature are concerned about trajectory ensembles conditioned on simple time-integrated observables, it shall be interesting to explore complex quantities, e.g., stochastic efficiency.

\begin{acknowledgments}
This work was supported by the National Science Foundation of China under Grants No. 12075016 and No. 11575016.
\end{acknowledgments}

\appendix

\section{Explanation of Eqs~(\ref{longtimedensitymatrix}) and~(\ref{scaledcumulantgeneratingfunction}) }
The moment generating function~(\ref{heatmomentgenreatingfunction}) can be rewritten as $
\Phi_T (\chi)=\sum_{m,n=1}^N  p_m\phi_{m|n}(\chi,T)$,
where
\begin{eqnarray}
\label{phim1n}
 \phi_{m|n}(\chi,T)=\sumint_{M=0}^\infty {\cal D}(t)\hspace{0.1cm}  p_{m|n}({\chi},\overrightarrow{\omega}_M,T) e^{\chi Q_T(\overrightarrow{\omega})} =\left.\langle u_n(t)|\hat \rho(t)|u_n(t)\rangle\right|_{t=T}.
\end{eqnarray}
Obviously, $\phi_{m|n}(\chi,T)$ is also the moment generating function of the stochastic heat current for the special quantum trajectory ensemble, in which the initial and terminal Floquet bases are constrained at $| u_m(0)\rangle$ and $|u_n(T)\rangle$, respectively. In the second equation of Eq.~(\ref{phim1n}), the operator $\hat\rho(t)$ ($0\le t\le T$) satisfies
\begin{eqnarray}
\label{equationofmotionforrhohat}
\partial_t \hat{\rho}(t)&=&{\cal L}_\chi(t)[\hat{\rho}(t)],
\end{eqnarray}
and its initial condition is $|u_m(0)\rangle\langle u_m(0)| $. To derive this result, we use the quantum trajectory probability formula of Eq.~(\ref{probabilityformofdensitymatrix}) and the ensemble average of Eq.~(\ref{probdensityfunctional})~\cite{Liu2016a,Liu2018,Liu2020}. Eq.~(\ref{equationofmotionforrhohat}) has been named the modified quantum master equation~\cite{Esposito2009,Garrahan2010,Gasparinetti2014,Cuetara2015} or tilted quantum master equation~\cite{Carollo2018}, where another technique, counting field statistics, is used in the derivation. If we express the evolution equation in the Floquet basis, we have
\begin{eqnarray}
\frac{d}{dt} \overrightarrow \phi_{m}(\chi,t)=\textbf{R}(\chi) \overrightarrow \phi_{m}(\chi,t),
\end{eqnarray}
where $ \overrightarrow \phi_{m}$ is the transpose of vector $(\phi_{m|1},\cdots,\phi_{m|N})$ and its initial condition is $  \phi_{m|n}(0)=\delta_{m,n}$, $n=1,\cdots,N$. The non-diagonal matrix element of ${\bf R}(\chi)$ is
\begin{eqnarray}
&&{R}_{ij}(\chi) =\sum_{\omega\neq q\Omega} e^{\chi \hbar\omega} r(\omega)|\langle u_i(t)|A(\omega,t)|u_j(t)\rangle|^2, 
\end{eqnarray}
($i\neq j$), and the diagonal elements are
\begin{eqnarray}
{R}_{ii}(\chi)= \sum_{\omega=q\Omega}  e^{\chi\hbar\omega}  r(\omega)|\langle u_i(t)|A(\omega,t)|u_i(t)\rangle|^2-
\sum_{\omega} r(\omega) \sum_{j'} |\langle u_{j'}(t)|A(\omega,t)|u_i(t)\rangle|^2.
\end{eqnarray}
$i,j,j'=1,\cdots N$. We emphasize that ${\bf R}(\chi)$ is a constant matrix. Then, we formally express the moment generating functions as
\begin{eqnarray}
\phi_{m|n}(\chi,T)=\left(e^{T{\bf R}(\chi)}\right)_{mn}.
\end{eqnarray}
Because these moment generating functions are always positive, according to the Perron-Frobenius theorem, we immediately arrive at the conclusion that the matrix ${\bf R}(\chi)$ has a unique maximum positive eigenvalue with a positive left eigenvector, i.e., Eq.~(\ref{lefteigenvector}). When the duration $T$ tends to infinity, Eq.~(\ref{scaledcumulantgeneratingfunction}) is proven. There is a special case, $\chi=0$. Under this situation, the maximum left eigenvector is a unit vector, $(1,\cdots,1)$, and the maximum eigenvalue trivially equals $0$. Note that the corresponding right eigenvector is the stationary probability distribution, $(P_1,\cdots,P_N)$. Hence, Eq.~(\ref{longtimedensitymatrix}) is proven as well.

\section{$\Lambda'(\chi')$ of the auxiliary open quantum system }
The moment generating function $\Phi_T'(\chi')$ of the auxiliary open quantum system can be solved as that of the original quantum system: $\Phi_T'(\chi')={\rm Tr}[\hat\rho'(T)]$ and 
$\partial_t \hat{\rho}'(t)={\cal L'}_{\chi'}(t)[\hat{\rho}'(t)]$, where the generator of the tilted quantum master equation is
\begin{eqnarray}
{\cal L}'_{\chi'}(t)(O)&=&-\frac{i}{\hbar}[H(t), O]+\sum_{\omega} r(\omega )\left[e^{\chi'\hbar\omega}A'(\omega,t)O A'^\dag(\omega ,t)- \frac{1}{2}\left\{A'^\dag (\omega ,t)A'(\omega ,t),O\right\}\right].
\end{eqnarray}
Then, we construct the matrix ${\textbf R}'(\chi')$ and its elements as follows.
The non-diagonal elements are
\begin{eqnarray}
\label{generalij}
{ R}'_{ij}(\chi')&=&\frac{l_i^2(\chi)}{l_j^2(\chi)}\sum_{\omega\neq q\Omega} e^{(\chi'+\chi) \hbar\omega} r(\omega)|\langle u_i(t)|A(\omega,t)|u_j(t)\rangle|^2  \nonumber \\
&=&\frac{l_i^2(\chi)}{l_j^2(\chi)} { R}_{ij}(\chi'+\chi),
\end{eqnarray}
($i\neq j$), and the diagonal elements are
\begin{eqnarray}
\label{generalii}
{R}'_{ii}(\chi')&=& \sum_{\omega=q\Omega}  e^{(\chi'+\chi)\hbar\omega}  r(\omega)|\langle u_i(t)|A(\omega,t)|u_i(t)\rangle|^2-
\sum_{\omega} e^{\chi\hbar\omega}r(\omega) \sum_{j'} \left(\frac{l_{j'}}{l_i}\right)^2|\langle u_{j'}(t)|A(\omega,t)|u_i(t)\rangle|^2,\nonumber \\
&=&{R}_{ii}(\chi'+\chi)-\Lambda(\chi).
\end{eqnarray}
Eq.~(\ref{lefteigenvector}) is used here. We immediately find that the vector $\overrightarrow {l'^2}(\chi')=(l'^2_1(\chi'),\cdots,l'^2_N(\chi'))$ with elements
\begin{eqnarray}
l'^2_n(\chi')=\frac{l_n^2(\chi'+\chi)}{l_n^2(\chi)},
\end{eqnarray}
$n=1,\cdots,N$, is the eigenvector of Eq.~(\ref{lefteigenvectorauxiliarysystem}), while the eigenvalue is Eq.~(\ref{SCGFauxiliarysystem}).

We shall mention that $\Lambda'(\chi')$ can be alternatively obtained by abstract operator calculations as Carollo et al.~\cite{Carollo2018} did. First, Eq.~(\ref{lefteigenvectorauxiliarysystem}) has an equivalent operator expression:
\begin{eqnarray}
\label{eigenvalueauxiliarysystem}
\partial_t  l'^2_{\chi'}(t)+ {\cal L}'^*_{\chi'}(t)[l'^2_{\chi'}(t)]=\Lambda'(\chi') l_{\chi'}'^2(t),
\end{eqnarray}
where
\begin{eqnarray}
{\cal L'}_{\chi'}^*(t)(O)&=&\frac{i}{\hbar}[H(t), O]+\sum_{\omega} r(\omega )\left[e^{\chi'\hbar\omega}A'^\dag(\omega ,t)O A'(\omega ,t)- \frac{1}{2}
\left\{A'^\dag (\omega ,t)A'(\omega ,t),O\right\}\right].\nonumber \\
&=&l^{-1}_\chi(t) {\cal L}^*_{\chi'+\chi}(t)\left[ l_\chi(t)Ol_\chi(t) \right] l^{-1}_\chi(t)-\Lambda(\chi)O + l_\chi^{-1}(t) (\partial_t l_\chi(t))O+O(\partial_t l_\chi(t))l_\chi^{-1}(t).
\end{eqnarray}
Then, it is direct to verify that the operator
\begin{eqnarray}
l'^2_{\chi'}(t)&=&\sum_{n=1}^N l'^2_n(\chi')|u_n(t)\rangle \langle u_n(t)|=l^{-1}_\chi(t) l^2_{\chi'+\chi}(t) l^{-1}_\chi(t)
\end{eqnarray}
satisfies Eq.~(\ref{eigenvalueauxiliarysystem}) and that $\Lambda'(\chi')$ is simply Eq.~(\ref{SCGFauxiliarysystem}).

\end{document}